\documentclass[final,5p,times,twocolumn]{elsarticle}

\usepackage{amssymb}
\usepackage{amsmath}
\usepackage{epsfig,latexsym}
\usepackage{bm}
\usepackage{multirow}
\usepackage[dvips]{color} 

\newcommand{\Psfig}[2]{\includegraphics[width=#1]{Figs/#2}}
\newcommand{\Msun}{M_{\odot}}
\newcommand{\muB}{\mu_{B}}

\newcommand{\mud}{\mu_{d}}
\newcommand{\rhoB}{\rho_{B}}
\newcommand{\MeV}{\mathrm{MeV}}
\newcommand{\fm}{\mathrm{fm}}
\newcommand{\PNJLx}{P-NJL$_8$}
\newcommand{\TCP}{T_\mathrm{CP}}
\newcommand{\muCP}{\mu_\mathrm{CP}}

\begin{document}

\begin{frontmatter}



\title{Possibility of QCD critical point sweep \\
during black hole formation
}



\author[YITP]{A. Ohnishi}
\author[Physics]{H. Ueda}
\author[YITP,Physics]{T. Z. Nakano}
\author[YITP]{M. Ruggieri}
\author[Numazu]{K. Sumiyoshi}

\address[YITP]{Yukawa Institute for Theoretical Physics, Kyoto University,
Kyoto 606-8502, Japan}

\address[Physics]{Department of Physics, Faculty of Science, Kyoto University,
Kyoto 606-8502, Japan}

\address[Numazu]{Numazu College of Technology, Ooka 3600, Numazu,
Shizuoka 410-8501, Japan}

\begin{abstract}
We discuss the possibility to probe the QCD critical point 
during the dynamical black hole formation from a gravitational collapse
of a massive star,
where the temperature and the baryon chemical potential
become as high as $T \sim 90~\MeV$ and $\muB \sim 1300~\MeV$.
Comparison with the phase boundary in chiral effective models
suggests that quark matter is likely to be formed before the horizon is formed.
Furthermore, the QCD critical point may be probed
during the black hole formation.
The critical point is found to move in the lower temperature direction
in asymmetric nuclear matter,
and in some of the chiral models
it is found to be in the reachable region
during the black hole formation processes.
\end{abstract}


\end{frontmatter}


\section{Introduction}
\label{Sec:Intro}
The critical point (CP) of the Quantum Chromodynamics
(QCD)~\cite{Asakawa:1989bq}
may be regarded as a corner stone of the QCD phase diagram;
the cross over~\cite{Aoki:2006we} and the first order phase boundaries
between the hadron and quark-gluon phases are connected by CP,
then it determines the global structure of the phase diagram.
Against its importance,
the existence, number and location of CP are not yet established
in theoretical calculations.
The sign problem makes the lattice Monte Carlo simulation difficult
in the large baryon-chemical potential ($\muB$)
region~\cite{LQCD-CEP},
the strong coupling expansion of lattice QCD~\cite{SC-LQCD}
is not yet reliable to describe the real world,
and the predictions in effective models
are spread in the $T-\muB$ plane~\cite{Stephanov:2007fk}.
Thus further experimental and theoretical developments
are necessary to reveal the properties of CP.
The search for CP in heavy ion collisions is
ongoing in the beam energy scan
programs at RHIC~\cite{BES},
and CP is one of the most important targets 
in the forthcoming FAIR facility.
The most characteristic feature of CP is
the divergence of the coherence length $\xi$.
The phase transition becomes the second order,
and large fluctuations of the order parameters are expected
in a volume of the size $\xi^3$.
On the basis of this idea,
various signatures of CP have been suggested theoretically~\cite{Sig-CEP}.
However, since the system size and the evolution time are limited,
it is not an easy task to observe the divergence signature of $\xi$
in heavy-ion collisions~\cite{Stephanov:1999zu}.
In addition, if the baryon chemical potential of CP
($\muCP$) is above 500 MeV,
CP may not be reachable
in heavy-ion collisions.
Therefore, it is important to examine other candidate sites 
where hot and dense matter is formed and CP is reachable.

A gravitational collapse of a massive star may be one of the promising
candidates for the CP hunting.
It has been argued that the transition to quark matter might trigger
the second collapse and bang in supernova
explosions~\cite{Hatsuda:1987ck,Sagert:2008ka}.
Recent calculation suggests that the QCD phase transition may take place
during the core-collapse of a star with $M=(10-15) \Msun$
when one uses the relativistic equation of state
(Shen EOS)~\cite{Shen-EOS} combined with the bag model EOS
at high $T$ or $\rhoB$ for a small bag constant,
$B^{1/4}\simeq 162~\MeV$~\cite{Sagert:2008ka}.
The QCD phase transition leads to the second shock,
and is suggested to give successful supernova explosion
even in a simulation with spherical symmetry.
Non-rotating massive stars with mass $M \gtrsim 20 \Msun$
are expected to collapse without supernova explosions
and to form a black hole (BH)~\cite{StarFate}.
In Refs.~\cite{Sumiyoshi:2006id,Sumiyoshi:2007pp,Sumiyoshi:2008kw},
the BH formation processes are calculated
by using the neutrino-radiation hydrodynamical simulations
in general relativity.
In the collapse and bounce stage of a 40$\Msun$ star,
the core bounce launches the shock wave, but the shock wave stalls
due to the collisions with falling matter (accretion)
and goes down to the surface of the compact object.
The proto-neutron star is born at center and gradually contracts.
Because of the accretion, the proto-neutron star mass increases rapidly
and reaches the critical mass.
The dynamical collapse occurs again at this point and the BH is formed
at 1.3 s after the bounce in the case of the Shen EOS~\cite{Shen-EOS}.
If we adopt a relativistic EOS with hyperons (Ishizuka EOS)~\cite{IOTSY}
or the Lattimer-Swesty EOS~\cite{Lattimer:1991nc},
the second collapse becomes more rapid ($\sim$ 0.7 s after the bounce),
while the average neutrino energies are lower 
with Ishizuka EOS~\cite{Sumiyoshi:2008kw}.
The combination of the neutrino duration time and the neutrino energy
may be used as a signal of the hyperon emergence or other new degrees
of freedom during the BH formation~\cite{Nakazato}.

During the BH formation,
hot $(T \sim 90~\MeV)$ and dense $(\rhoB \sim 4 \rho_0)$ matter is created.
The temperature and density in BH formation are significantly higher than those
in the model explosion calculation of supernova,
where the highest temperature and density are
$(T,\rhoB) \sim (21.5~\MeV, 0.24~\fm^{-3})$~\cite{IOTSY}.
In hotter and denser environment during BH formation
compared with the supernova explosions,
we have a larger possibility of creating a new form of matter,
such as the dense quark matter.
Thermodynamical variables at a given time vary as a function of radius
in a proto-neutron star and form a line in the $T-\muB$ plane.
This thermodynamical line,
referred to as the BH formation profile in the later discussions,
evolves with time and may pass through CP and the vicinity.
We call here this situation as the {\em CP sweep}.

In this Letter, we examine the location of the QCD phase boundary and
CP in two-flavor chiral effective models
at finite isospin chemical potential $\delta\mu \equiv (\mu_d-\mu_u)/2$,
and discuss the possibility of the CP sweep during the BH formation
from a gravitational collapse of a massive star.
First, we compute the CP location 
in the $T-\muB$ plane in chiral effective models
such as the Nambu-Jona Lasinio (NJL) model~\cite{NJL},
the Polyakov loop extended Nambu-Jona-Lasino (P-NJL)
model~\cite{Meisinger:1995ih,Fukushima:2003fw,PNJL},
P-NJL model with eight-quark interaction~\cite{Kashiwa:2007hw},
and the Polyakov loop extended quark-meson (PQM)
model~\cite{Schaefer:2007pw,Skokov:2010sf}.
The bag model EOS adopted in Ref.~\cite{Sagert:2008ka} is not suited
to the present purpose, since it does not have CP.
In the dynamical BH formation,
we have abundant neutrinos, and approximate
$\beta$ equilibrium including neutrinos are realized inside the neutrino sphere,
$\mu_n-\mu_p=\mu_e-\mu_\nu$,
while neutrinos are out of equilibrium outside.
In both cases, it is necessary to take account of
finite isospin chemical potential $\delta\mu$
as another independent thermodynamical
variable~\cite{IsoMu},
rather than imposing the neutrino-less $\beta$ equilibrium condition
($\delta\mu=\mu_e/2$)
in order to examine the CP property during the BH formation.
Recently, P-NJL model with isospin chemical potential has been investigated
with~\cite{Abuki}
and without~\cite{Sasaki:2010jz} 
neutrino-less $\beta$ equilibrium conditions.
At finite $\delta\mu$, we naively expect that
CP moves in the lower $T$ direction
because of the larger $d$-quark chemical potential $\mud=\muB/3 + \delta\mu$.
Since the matter passes through the high $\muB$ and low $T$ region,
the reduction of the CP temperature $\TCP$ is essential
for the CP sweep during the BH formation.
Next, we compare the results of
the CP location in the chiral effective models
with the evolution of thermodynamical variables $(T,\muB,\delta\mu)$
during the BH formation obtained in Ref.~\cite{Sumiyoshi:2006id}.
It should be noted that we compare
the results of the CP location in chiral effective models
and the thermodynamical condition $(T,\muB)$ calculated with the hadronic EOS.
This comparison is relevant,
since the thermal trajectory should be the same
even if we use the combined EOS of quark and hadronic matter,
as long as the hadronic EOS is reproduced at low $T$ and $\muB$
in the combined EOS.
Finally, we discuss the possibility of the CP sweep during the BH formation
from the above comparison.

\section{Polyakov loop extended chiral effective models}
In this section, we summarize the chiral effective models,
the NJL, P-NJL, and PQM models,
which we use in computing the CP location in the $T-\muB$ plane.

\subsection{NJL model}
The Lagrangian density of the two flavor NJL model is given by
\begin{align}
&{\cal L}_\mathrm{NJL} = \bar q\left(i\gamma^\mu\partial_\mu - m_0\right) q 
+G_\sigma\left[(\bar q q)^2 + (\bar q i\gamma_5 \bm\tau q)^2\right]
\nonumber \\
&- G_\rho\left[(\bar q \gamma^\mu \bm \tau q)^2 + (\bar q i\gamma_5 \gamma^\mu \bm \tau q)^2\right]
    - G_\omega\left[(\bar q \gamma^\mu q)^2 + (\bar q i\gamma_5 \gamma^\mu  q)^2\right] \ ,
    \label{eq:Lagr}
\end{align}
where $q$ denotes a quark field with Dirac, color and flavor indices;
$\bm\tau$ is the Pauli matrices in the flavor space.
In what follows, we take $G_\rho = G_\omega \equiv G_V$,
which amounts to take $\omega$ and $\bm\rho$ mesons degenerate in the vacuum.
We can fix the value of $G_S$ by fitting
the well known properties of the QCD vacuum.
We will take the ratio $G_V/G_S$ as a free parameter.

We are interested in unbalanced populations of $u$ and $d$ quarks,
not in the neutrino-less $\beta$-equilibrium.
For this reason, we introduce two independent chemical potentials
for $u$ and $d$ quarks.
At the mean field level, the effect of the vector interaction
is to shift the quark chemical potentials:
the flavor singlet interaction gives a contribution proportional to
$\rho_u + \rho_d$; on the other hand, the flavor triplet
interaction gives a contribution proportional to the isospin
density, $\rho_u - \rho_d$. Keeping this into account,
and for later convenience, we define
\begin{equation}
\tilde{\mu}_u = \mu - \delta\mu -4 G_V \rho_u~,
~~~\tilde{\mu}_d = \mu + \delta\mu -4 G_V \rho_d~,
\label{eq:chemPot}
\end{equation}
where $\mu$ and $\delta\mu$ represent
chemical potentials conjugated to the total quark number density
and to isospin density, respectively.

The one-loop thermodynamic potential can be represented as the sum of
the the vacuum $\Omega_0$
and the thermal (finite temperature and finite chemical) $\Omega_T$
contributions,
\begin{align}
\Omega_\mathrm{NJL} =& \Omega_0 + \Omega_T , \label{eq:Otot}
\\
\Omega_0
=& \frac{\Sigma^2}{4 G_\sigma}
  - 2 N_c\sum_{f}\int\frac{d^3\bm p}{(2\pi)^3} F(\bm p^2,\Lambda) E_p
\ ,\label{eq:Ovac}
\\
\Omega_T =
&- 2 T N_c\sum_{f}\int\frac{d^3\bm p}{(2\pi)^3}
\log\left(1 + e^{-\beta {\cal E}_+^f}\right)
    \left(1 + e^{-\beta {\cal E}_-^f}\right)
\nonumber\\
&- 2 G_V\left(\rho_u^2 + \rho_d^2\right)
,\label{eq:Otemp}
\\
{\cal E}_{\pm}^f =& E_p \pm \tilde{\mu}_f~
\ , \label{eq:calEdef}
\end{align}
where
$E_p=\sqrt{\bm p^2 + M^2}$ with $M = m_0 + \Sigma$,
and $\Sigma \equiv -2G_\sigma\langle\bar q q\rangle$
corresponds to the mean field quark self-energy.
In Eq.~\eqref{eq:Ovac}, we have treated the ultraviolet divergence
of the vacuum energy by the use of a smooth regulating function,
$F(\bm p^2,\Lambda) = [1+(\bm p^2)^5/\Lambda^{10}]^{-1}$.
This is done just for numerical convenience; quantitatively, we
have checked that the results are consistent with the more common
hard cutoff regularization scheme, within a few percent.
The thermal part $\Omega_T$ is a finite contribution
which does not need any regularization.

\subsection{The P-NJL model}
The P-NJL model Lagrangian density is still specified by
Eq.~\eqref{eq:Lagr}, with the derivative replaced by a covariant
one: $\partial_\mu \rightarrow D_\mu = \partial_\mu - i A_\mu$.
Here, $A_\mu$ is a temporal, static and homogeneous background gluon field
related to the Polyakov loop $P$, whose expectation value is computed
self-consistently.
The one-loop thermodynamic potential in Eq.~\eqref{eq:Otot} is replaced
with~\cite{Fukushima:2003fw}
\begin{align}
\Omega_\mathrm{P-NJL} =& \Omega_0 + \Omega_T + {\cal U}(P,\bar{P},T) \ ,
\label{eq:OtotP}
\\
  \mathcal{U}[P,\bar P,T] =& T^4\biggl\{-\frac{a(T)}{2}
  \bar P P + b(T)\ln H(P,\bar{P})
 \biggr\} \;,
\label{eq:Poly}
\\
P =& \frac{1}{N_c}\text{Tr}\left[{\cal P}\exp\left(i\int_0^\beta
d\tau A_4\right)\right] \equiv \frac{1}{3}\text{Tr} e^{i\phi/T}~,
\label{eq:PolDef}
\end{align}
where ${\cal U}(P,\bar{P},T)$ is
the Polyakov loop effective potential~\cite{PNJL}
with
$H(P,\bar{P})= 1-6\bar PP + 4(\bar P^3 + P^3) -3(\bar PP)^2$,
$a(T) = a_0 + a_1 (T_0/T)+ a_2 (T_0/T)^2$,
and
$b(T) = b_3(T_0/T)^3$.
We adopt the Polyakov gauge, where $\phi$ is specified by
$\phi = \phi_3 \lambda_3 + \phi_8 \lambda_8$.
The standard choice of the parameters reads~\cite{PNJL}
$a_0 = 3.51\,, \quad a_1 = -2.47\,, \quad a_2 = 15.2\,, \quad b_3 = -1.75$.
The parameter $T_0$ in Eq.~\eqref{eq:Poly} sets the deconfinement
scale in the pure gauge theory, i.e. $T_c = 270$ MeV.

The thermal part of the thermodynamic potential
is now given by~\cite{Fukushima:2003fw}
\begin{align}
\Omega_T =& - 2 G_V\left(\rho_u^2 + \rho_d^2\right)  - 2 T
\sum_{f}\int\frac{d^3\bm p}{(2\pi)^3} \log
\left(F_+^f F_-^f\right)~,\label{eq:OtempP}
\\
F_+^f =& 1+3\bar P e^{-\beta {\cal E}_+^f } + 3Pe^{-2\beta{\cal
E}_+^f} +e^{-3\beta{\cal E}_+^f}~, \label{eq:FPdef}\\
F_-^f =& 1+3 P e^{-\beta {\cal E}_-^f } + 3 \bar P e^{-2\beta{\cal
E}_-^f} +e^{-3\beta{\cal E}_-^f}~.\label{eq:FMdef}
\end{align}
In Eq.~\eqref{eq:FPdef}, the addenda on the r.h.s. correspond to
the thermal contribution of zero, one, two and three quark states,
respectively. Analogously, Eq.~\eqref{eq:FMdef} is
the thermal contribution of antiquarks.

\subsection{P-NJL model with eight-quark interaction (\PNJLx)}
With eight-quark interaction, the quark Lagrangian is
~\cite{Kashiwa:2007hw}
\begin{align}
{\cal L} =& {\cal L}_\mathrm{NJL}
+ G_{\sigma 8}\left[(\bar q q)^2 + (\bar q i\gamma_5 \bm\tau q)^2\right]^2
\ .
    \label{eq:Lagr8}
\end{align}
We do not include here the eight-quark interaction in the vector channel.
The one loop thermodynamic potential is still given by
Eq.~\eqref{eq:OtotP}, with
\begin{equation}
\Omega_0 = \frac{3G_{\sigma 8}}{16 G_{\sigma}^4}\Sigma^4 +
\frac{\Sigma^2}{4G_{\sigma}} - 2 N_c\sum_{f}\int\frac{d^3\bm
p}{(2\pi)^3} F(\bm p^2,\Lambda) E_p~,\label{eq:Ovac8}
\end{equation}
and $E_p = \sqrt{\bm p^2 + M^2}$, with $M = m_0 + \Sigma +\Sigma^3
(G_{\sigma 8}/2G_\sigma^3)$.
For simplicity, we call this model P-NJL$_8$.

P-NJL$_8$ parameters ($\Lambda, G_\sigma, G_{\sigma 8}, m_0$) are determined
to fix $f_\pi$, $m_\pi$, $m_\sigma$, and $\langle{\bar{u}u}\rangle$,
and as a result, the chiral and deconfinement transition temperatures
are found to roughly agree~\cite{Kashiwa:2007hw}.
The eight-quark interaction is found to make chiral cross over sharper
at $\mu=0$;
$\mu$ makes this cross over stronger, and eventually leads to the first order
transition. Hence eight-quark interaction causes chemical potential of CP
to be smaller.

\subsection{PQM model}
The Lagrangian density of the PQM model is given by
\begin{align}
{\cal L} =& \bar{q}\left[
i\gamma^\mu D_\mu
- g(\sigma + i\gamma_5\bm\tau\cdot\bm\pi)
- g_v\gamma^\mu(\omega_\mu + \bm\tau\cdot\bm{R}_\mu)
\right]q 
\nonumber\\
+& \frac{1}{2}(\partial_\mu\sigma)^2 +
\frac{1}{2}(\partial_\mu\bm\pi)^2 -U(\sigma,\bm\pi)
-{\cal U}(P,\bar P,T)
\nonumber\\
-& \frac{1}{4} \omega_{\mu\nu}\omega^{\mu\nu}
- \frac{1}{4} \bm R_{\mu\nu} \cdot \bm R^{\mu\nu}
+ \frac{1}{2} m_v^2 (\omega_\mu\omega^\mu
	+ \bm R_\mu\cdot \bm R^\mu)
~.
\end{align}
The mesonic potential is
$U(\sigma,\bm\pi)=\lambda\left(\sigma^2+\bm\pi^2-v^2\right)^2/4-h\sigma$,
and $\omega_{\mu\nu}$ and $\bm R_{\mu\nu}$ are the field tensors of
the $\omega$ and $\rho$ mesons.
We use the same Polyakov loop effective potential
as that in the P-NJL model, Eq.~\eqref{eq:Poly}.
In the mean field approximation, the thermodynamic potential is
given by
\begin{align}
\Omega_\mathrm{PQM} =& {\cal U}(P,\bar P,T) + U(\sigma,\bm\pi=0) +\Omega_0 +
\Omega_T~,\\
\Omega_0 =& -2N_f N_c\int\frac{d^3\bm p}{(2\pi)^3} E_p
\theta(\Lambda^2 -\bm p^2)~,
\end{align}
where $\Omega_0$
corresponds to the regularized fermion vacuum energy
with $M=g\sigma$, and
$\Omega_T$ is still given by Eq.~\eqref{eq:OtempP}
with $G_V = g_v^2/2m_v^2$.
While the PQM model is renormalizable and an elegant procedure of
dimensional renormalization is feasible~\cite{Skokov:2010sf},
it is enough to cut large momenta by a hard cutoff for our purposes.

\subsection{Model parameterization}
In this study, we fix $\mu$ and $\delta\mu$, and compute
$\rho_u$, $\rho_d$, the chiral condensate
and the Polyakov loop expectation values self-consistently,
requiring the stationary condition of the thermodynamic potential.
In the case of the NJL model and the P-NJL model without eight-quark
interaction, the parameters $G_\sigma$, $\Lambda$ and $m_0$ are
chosen in order to reproduce the QCD vacuum properties
$\langle\bar u u\rangle = (-250~\text{MeV})^3$,
$f_\pi = 92.4$ MeV and $m_\pi = 139$ MeV.
They are given as
$\Lambda = 618.98~\MeV$, 
$G_\sigma = 2.05/\Lambda^2$, 
and
$m_0 = 5.28~\text{MeV}$.
The parameter $T_0$ in the Polyakov loop effective potential is taken to be
$T_0 = 210$ MeV. With this parameter choice,
the constituent quark mass in the vacuum is $M \approx 340$ MeV.
In \PNJLx, we use the parameterization in~\cite{Kashiwa:2007hw},
$\Lambda = 631.5~\MeV$,
$G_\sigma = 1.864/\Lambda^2$,
$G_{\sigma 8} = 11.435/\Lambda^8$,
and
$m_0 = 5.5~\MeV$,
which give the vacuum constituent quark mass $M \approx 353$ MeV.
For $G_V$, estimates exist based on
perturbative one gluon exchange~\cite{Hatsuda:1985ey},
$r\equiv G_V/G_\sigma = 0.5$;
on instanton-anti-instanton molecule model~\cite{Kitazawa:2002bc}, $r=0.25$;
and an interpolation is obtained by a fit of the Lattice data with the
\PNJLx\ model in~\cite{Sakai:2009dv}, $r=1$.
We here treat $r$ as a free parameter,
and compare the results with $r=0$ and $r=0.2$.

The parameters of the PQM model, $v$, $\lambda$, $g$ and $h$,
are fixed to reproduce some vacuum properties:
the chiral condensate in the vacuum,
$\sigma=f_\pi= 92.4~\MeV$;
the vacuum pion mass, $m_\pi^2 = h/f_\pi = (139\text{MeV})^2$;
the constituent vacuum quark mass, $M = g f_\pi = 335\text{MeV}$;
the sigma mass, given by 
$m_\sigma^2 = \partial^2\Omega/\partial\sigma^2 = (700~\MeV)^2 $.
In this article, we use the following parameter set:
$\Lambda = 600\text{MeV}$,
$v^2 = -(617.68\text{MeV})^2$,
and
$\lambda = 2.7255$.
We use the same Polyakov loop effective potential as that in the P-NJL model.
The vector meson mass is chosen to be $m_v=770~\MeV$.
The vector coupling is treated as a free parameter,
and we compare the results with $r=g_v/g=0$ and $0.2$ in the later discussions.

P-NJL models with SU(3)$_f$ are known to support hybrid neutron star mass of
1.74-2.12 $M_\odot$ when combined with a hadronic EOS
at low densities~\cite{Blaschke}.
By comparison, an SU(3)$_f$ version of PQM model connected with a hadronic EOS
can support the hybrid neutron star mass of
$2.0 M_\odot$~\cite{Dexheimer:2009va}.
Generally speaking, the EOS becomes softer with larger degrees of freedom,
then SU(2)$_f$ versions of these models are expected to give stiffer EOS
than in SU(3)$_f$.
Therefore SU(2)$_f$ chiral effective models adopted in this work
would result in larger maximum masses of neutron stars,
which are consistent with the recently observed
$1.97 \pm 0.04 M_\odot$ neutron star~\cite{Demorest:2010bx}.


\section{Critical point location and its sweeping}

\begin{figure}[hbt]
\begin{center}
\Psfig{7cm}{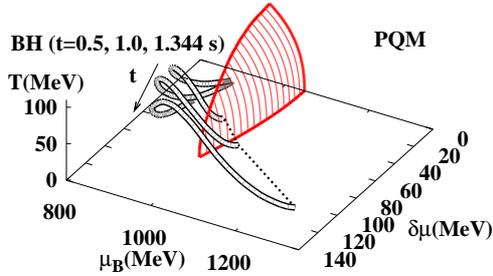}
\end{center}
\caption{
Phase diagram in $(T, \muB, \delta\mu)$ space.
First order phase boundaries $(T, \muB)$ calculated with PQM are shown
for several values of the isospin chemical potential, $\delta\mu$.
We also show the BH formation profile, 
(thermodynamical profile $(T,\muB,\delta\mu)$ during the BH formation)
at $t=0.5, 1.0, 1.344$ sec after the bounce
(double lines).
}\label{Fig:3D}
\end{figure}

\begin{table}
\caption{Location of CP, the transition chemical potential at $T=0$ ($\mu_c$),
and the type of the transition to quark matter during the BH formation.
All $T$ and $\mu$ values are given in the unit of MeV.
}
\label{table:CP}
\begin{center}
\begin{tabular}{l|c|rrrr|c}
\hline
\hline
Model &$r$& $\delta\mu$	& $\TCP$&$\muCP$& $\mu_c$ & BH \\
\hline
\multirow{5}{*}{NJL}
	& \multirow{3}{*}{0}	
		& 0	& 50	& 993	& 1095	& \multirow{3}{*}{CP sweep}\\
	 &	& 50	& 45	& 999	& 1065	\\
	 &	& 65	& 37	& 1005	& 1035	\\
\cline{2-7}
	& \multirow{2}{*}{0.2}
		& 0	& 22	& 1095	& 1110	& \multirow{2}{*}{Cross over}\\
	 &	& 50	& 10	& 1073	& 1074	\\
\hline
\multirow{5}{*}{P-NJL}
	&\multirow{3}{*}{0}
	 	& 0	& 106	& 975	& 1095	& \multirow{3}{*}{CP sweep}\\
	&	& 50	& 92	& 990	& 1065	\\
	&	& 65	& 86	& 996	& 1035	\\
\cline{2-7}
	&\multirow{2}{*}{0.2}
		& 0	& 74	& 1062	& 1110	& \multirow{2}{*}{Cross over}\\
	&	& 50	& 39	& 1068	& 1086	\\
\hline
\multirow{5}{*}{\PNJLx}
	&\multirow{3}{*}{0}
		& 0	& 145	& 600	& 1005	& \multirow{3}{*}{First order}\\
	&	& 50	& 125	& 678	&  900	\\
	&	& 65	& 118	& 690	&  870	\\
\cline{2-7}
	&\multirow{2}{*}{0.2}
		& 0	& 129	& 708	& 1020	& \multirow{2}{*}{First order}\\
	&	& 50	& 119	& 720	&  930	\\
\hline
\multirow{6}{*}{PQM}
	&\multirow{3}{*}{0}
		& 0	& 105	& 964	& 1046	& \multirow{3}{*}{CP sweep}\\
	&	& 50	& 87	& 979	& 1025	\\
	&	& 70	& 62	& 989	& 1007	\\
\cline{2-7}
	&\multirow{3}{*}{0.2}
		& 0	& 91	& 1006  & 1057	& \multirow{3}{*}{CP sweep}\\
	&	& 50	& 69	& 1016	& 1040	\\
	&	& 70	& 35	& 1020	& 1024	\\
\hline
\hline
\end{tabular}
\end{center}
\end{table}

\subsection{Critical point and phase boundary in asymmetric matter}

In Fig.~\ref{Fig:3D},
we show the isospin chemical potential dependence of
the first order phase boundary and the critical point in the PQM model.
We find a trend that the first order phase boundary shrinks
at finite isospin chemical potential.
Transition temperature at a given baryon chemical potential
$\muB=3\mu$ decreases,
and the transition chemical potential $\mu_c$ at $T=0$ also decreases.
We do not consider here the pion condensed phase,
because the $s$-wave pion condensation will not be realized
when we include the $s$-wave $\pi N$ repulsion~\cite{Ohnishi:2008ng}.
The CP location is sensitive to $\delta\mu$.
Compared with the results in symmetric matter,
$\TCP$ becomes smaller at finite $\delta\mu$
and reaches zero at $\delta\mu=\delta\mu_c \simeq (50-80)~\MeV$.
The downward shift of $\TCP$ may be understood from the density shift.
At low $T$ and without the vector interaction,
the quark density is proportional to $\mu^3$,
$\rho_{u,d}\propto (\mu \mp \delta\mu)^3$.
Then the sum of $u$ and $d$ quark density increases when $\delta\mu$ is finite,
and it simulates higher $\mu$,
where the transition temperature is lower.

In Table~\ref{table:CP},
we summarize the CP location $(T,\muB)$
for several values of $\delta\mu$ and $r=G_V/G_\sigma$
in NJL, P-NJL, \PNJLx, and PQM models.
In \PNJLx, our results at $r=0$ are in agreement
with those in Ref.~\cite{Sasaki:2010jz}.
The transition chemical potential at $T=0$ is 
in the range of $1000~\MeV < \mu_c < 1110~\MeV$.
$\mu_c$ is sensitive to the details of the interaction,
especially to the strength of the vector interaction.
The temporal component of the vector potential shifts the chemical potential
effectively as already introduced in Eq.~(\ref{eq:chemPot}).
In the momentum integral, we find the effective chemical potential
$\tilde{\mu}_f = \mu\mp\delta\mu - V_f$ appears,
where $V_f=4G_V\rho_f$ represents the vector potential for quarks.
The repulsive vector potential reduces the effects of the chemical
potential and consequently leads to an upward shift of $\mu_c$
by about 10-15 MeV at $r=0.2$.
When we increase the vector coupling from $r=0$ to $r=0.2$ in NJL,
the first order transition boundary is shifted upward in $\mu$
and $\TCP$ is reduced from 50 MeV to 22 MeV.
At larger vector coupling, the first order phase boundary disappears,
and the QCD phase transition becomes the cross over at any $\mu$.
This trend also applies to the P-NJL and PQM models;
the phase boundary is shifted in the larger $\mu$ direction
and shrinks in the $T$ direction with finite vector interaction.
In \PNJLx, the first order phase transition is robust
and survives with larger vector interaction such as $r=0.8$,
while the effects of the vector interaction is qualitatively the same.

\begin{figure*}[bth]
\begin{center}
\Psfig{18cm}{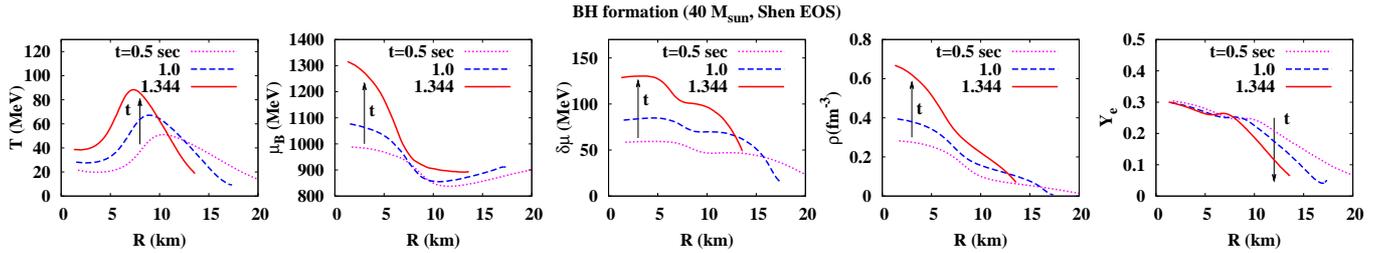}
\end{center}
\caption{
The BH formation profile, ($T, \muB, \delta\mu$), as a function of the radius.
We also show the baryon density ($\rhoB$) and the electron fraction
($Y_e \equiv \rho_e / \rhoB$).
Results are shown for the gravitational collapse of a 40 $\Msun$ star
at $t=$ 0.5 sec (dotted lines),
1.0 sec (dashed lines),
and 1.344 sec (solid lines, just before the horizon formation).
}\label{Fig:BH}
\end{figure*}

\subsection{BH formation profile}

In Fig. \ref{Fig:BH},
we show the BH formation profile
$(T,\muB,\delta\mu)$~\cite{Sumiyoshi:2006id}
calculated by using the Shen EOS
at $t=0.5, 1.0$ and $1.344$ sec after the bounce during the BH formation
from a 40 $\Msun$ star
in the proto-neutron star core,
where the mass coordinate from the center is $M < 1.6 M_\odot$.
The time $t=1.344$ sec is just before the horizon formation.
From the outer to the inner region of the proto-neutron star,
$T$ first increases from $T\sim 10~\MeV$ to $T\sim~(50-90)~\MeV$
in the middle heated region, and decreases again inside.
The central density grows from $\rhoB \sim \rho_0$ at bounce
to $2\rho_0$, $2.5\rho_0$ and $4\rho_0$ at 0.5, 1.0 and 1.344 sec,
respectively.
The charge to baryon ratio ($Y_e$) is
less than 0.3 inside the proto-neutron star~\cite{Sumiyoshi:2007pp}.
The isospin chemical potential is found to be $50-130$ MeV in the inner region.
The baryon chemical potential $\muB$ is found to go over 1300 MeV
in the central region just before the horizon formation at $t=1.344$ sec.

The maximum chemical potential is much larger than the $\Lambda(1115)$ mass,
and hyperons are expected to emerge.
Actually, hyperons are formed abundantly
when we use Ishizuka EOS including hyperons~\cite{Sumiyoshi:2008kw},
while the proto-neutron star collapses earlier
and the maximum $\muB$ ($\sim 1100$ MeV) is lower.

It should be noted that
the maximum mass of neutron stars is 2.17 $M_\odot$ in Shen EOS,
which does not contradict to the observation of 
$1.97 M_\odot$ neutron star~\cite{Demorest:2010bx}.
We need care in discussing the results of Ishizuka EOS,
since the maximum mass is $1.63 M_\odot$ and
it cannot support the $1.97 M_\odot$ mass neutron star.
It would be important to discuss the hyperon admixture with more repulsive
hyperon potentials at high densities, which support heavier neutron stars.

\begin{figure*}[tbhpf]
\begin{center}
\Psfig{10cm}{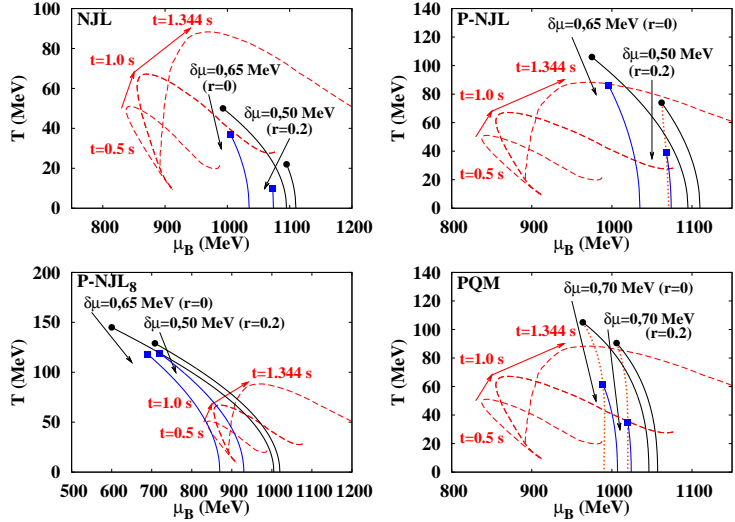}
\end{center}
\caption{
Critical point, phase boundary and the BH formation profile.
Critical point in symmetric (circles) and asymmetric (squares) matter
and the first order phase transition boundaries (solid lines)
in chiral effective models
are compared with
the BH formation profile $(T,\muB)$ at $t=0.5, 1.0$ and 1.344 sec
(dashed lines).
Top-left, top-right, bottom-left and bottom-right panels
show the results in the NJL, P-NJL, \PNJLx, and PQM models,
respectively.
We show the results without ($r=0$)  and with ($r>0$) vector interaction.
In P-NJL with vector interaction and PQM,
we also show the CP trajectory (dotted lines).
Arrows around the BH formation profile show time evolution,
and arrows around the critical points show the isospin chemical potential
dependence.
}\label{Fig:cepPDs}
\end{figure*}

\subsection{Possibility of critical point sweep}
\label{subsec:BH}

We shall now compare the CP location and the phase boundary
with the BH formation profile.
In Fig.~\ref{Fig:cepPDs},
we compare the phase boundaries and the CP location
in NJL (left-top), P-NJL (right-top), \PNJLx\ (left-bottom)
and PQM (right-bottom) models
with the BH formation profile, $(T, \muB)$.
As already mentioned,
the transition chemical potential is in the range of 
$1000~\MeV < \mu_c < 1110~\MeV$ in symmetric nuclear matter,
and it decreases at finite $\delta\mu$.
During the BH formation,
the baryon chemical potential reaches around 1000,  1100 and 1300 MeV
in the central region of the proto-neutron star at
$t=0.5, 1.0$ and $1.344$ sec, respectively.
This comparison suggests that quark matter would be formed
between $t=0.5$ and $1.0$ sec in the central region of the proto-neutron star
in most of the models considered here.

Since
the CP location has strong dependence on model and parameter,
there are three possible types in the transition to the quark matter
during the evolution of matter toward the BH formation;
the first order transition,
the cross over transition,
and the CP sweep.
In \PNJLx\ (with and without vector interaction),
$\TCP$ is relatively high even in asymmetric matter,
then the matter experiences the first order phase transition,
and CP is not reached.
In NJL and P-NJL with vector interaction,
$\TCP$ decreases in asymmetric matter,
and CP already disappears in the central region at $t=1.0$ sec,
where the isospin chemical potential is large,
$\delta\mu \sim 70~\MeV$.
In this case, the BH formation profiile evolves above CP,
and the cross over transition to quark matter will proceed
without going through the first order boundary.
In NJL and P-NJL models without vector interaction
and PQM with and without vector interaction,
the BH formation profile goes through CP from below,
as shown in the double line in Fig.~\ref{Fig:3D}.
CP in symmetric matter is above the BH formation profile
at $t=1.0~\mathrm{sec}$,
while CP in asymmetric matter is below the line at $t=1.344~\mathrm{sec}$.
Since the matter in the central region is highly asymmetric
($\delta\mu=(50-130)$ MeV) at $t=1.344~\mathrm{sec}$,
some part of the off-center BH forming matter would go through CP
between $t=1$ and 1.344 sec, {\em i.e.} CP is swept.

CP sweep will take place in the off-center inner core.
As can be seen in Fig.~\ref{Fig:BH}, the highest density and baryon
chemical potential is realized at center, while temperature becomes highest
off-center due to 
the shock wave.
In the PQM $(r=0)$ case, as a typical example of CP sweep,
the quark matter formation takes place first at center with $T<\TCP$,
the quark matter region grows,
and the CP sweep takes place when the boundary temperature reaches $\TCP$.
We find from Fig.~\ref{Fig:cepPDs}
that CP will be swept at a little inner region than the highest $T$ radius,
and this corresponds to $R=6-8~\mathrm{km}$ as seen in Fig. \ref{Fig:BH}.

It is also interesting to discuss
whether the condition in core-collapse supernovae 
may cross the phase boundaries to the quark matter phase.  
We examined the numerical result of the gravitational collapse and bounce 
from the massive star of 15M$_\odot$ with Shen-EOS \cite{sum05}.  
The density and temperature at the core bounce are 
$(\rhoB, T)=(0.2 \mathrm{fm}^{-3}, 15~\mathrm{MeV})$ at center.  
The baryon chemical potential and temperature do not reach 
the phase transition region in this case.  
They do not rise enough during the stage of the stalled shock wave 
for 200 ms after the core bounce.  
Therefore, it is not likely to have the CP sweep in dense matter 
around the core bounce for successful explosions.  
The situation is similar in the numerical result of adiabatic collapse 
and bounce of the 15M$_\odot$ star using Ishizuka' hyperonic EOS
as discussed in~\cite{IOTSY}.  
However, 
there are still possibilities to have the quark matter and the CP sweep
in the late stage of core-collapse supernovae.
In the thermal evolution of the proto-neutron star for $\sim$ 20 sec 
after the birth in supernovae, the density and temperature increase 
further due to the contraction during the thermal evolution.  
In fact, the condition may become closer to that of CP
in the long term evolution after the bounce \cite{sum05}.  
In the massive proto-neutron star cases~\cite{Pons}, 
hyperons and quarks appear in the late stage due to the extreme conditions, 
where they reach the phase boundaries.  
Hence, the CP sweep may be realized 
in the evolution toward the black hole formation in the very massive 
proto-neutron stars, while the explosion goes on.  
It is also possible to have high density and temperature 
in the transition region to quark matter
at the second bounce of the core collapse supernova
when one adopts an EOS having a low transition baryon chemical
potential~\cite{Sagert:2008ka}.

There are two comments in order.
(a) While we have compared the CP location and the BH formation profile
in $(T,\muB,\delta\mu)$,
the relation between the densities and chemical potentials is highly
model dependent and the baryon densities in the quark and hadronic models
considered here
do not generally coincide;
For example, in NJL and P-NJL, the baryon densities on CP
are found to be $\rhoB = 0.30$ and $0.29~\fm^{-3}$ at $\delta\mu=65~\MeV$,
respectively, while $\rhoB=0.36$ and $0.31~\fm^{-3}$ in Shen EOS
at the corresponding $(T,\rhoB,\delta\mu)$.
In PQM with $r=0$ and 0.2,
$\rhoB=0.25$ and $0.23~\fm^{-3}$ at $\delta\mu=60~\MeV$, respectively,
while $\rhoB=0.33$ and $0.34~\fm^{-3}$ in Shen EOS
on these points.
Thus the baryon densities in chiral effective and hadronic models
have $10-50 \%$ differences in density at CP,
and the results shown in this article would be modified
when we perform a consistent calculation using a hybrid supernova matter EOS
which smoothly connect quark and hadronic matter EOSs.
However, the trajectories span a wide region in $(T, \rhoB)$ 
which are in a range expected for the phase transition lines and CP, 
then we expect the CP sweep or at least the transition to quark matter
even in a consistent calculation of the black hole formation.
(b) The symmetry energy in RMF models has a strong dependence in $\rhoB$
and may overestimate the effects at high densities.
When one adopts a weak $\rhoB$ dependence of the symmetry energy
at high densities, smaller isospin chemical potential results.
In this case, the cross over transition during BH formation becomes less 
probable, since $\TCP$ is kept to be high at small $\delta\mu$.

\section{Summary}

In this Letter, we have discussed the possibility
of the QCD phase transition to quark matter
and the critical point (CP) sweep
during the dynamical black hole (BH) formation.
We have compared the phase boundary and CP in chiral effective models
with the BH formation profile, thermodynamical variables $(T,\muB)$
calculated in the neutrino-radiation hydrodynamics.
For this comparison, it is necessary to consider asymmetric matter
at finite isospin chemical potential, $\delta\mu=(\mu_n-\mu_p)/2 \not= 0$.
The isospin chemical potential is found to reduce the temperature of
the critical point $\TCP$,
then we have a larger possibility of the CP sweep or the cross over transition
to quark matter.

In the chiral effective models considered here,
with and without the vector interaction,
the transition chemical potential at $T=0$ is found to be in the range
of $\mu_c=(1000-1110)~\MeV$ in symmetric matter,
and $\mu_c$ decreases at finite $\delta\mu$.
We can compare these values with the highest baryon chemical potential
realized during the BH formation, $\muB=1300~\MeV$.
Then if the baryon chemical potential is larger than 
the QCD phase transition chemical potential,
quark matter will be formed during the BH formation.
In order to conclude, however, it is necessary to examine
with the EOS which includes both baryonic and quark degrees of freedom.
The CP location is sensitive to the models and parameters.
We have found that there are three types of possibilities
of the transition to quark matter in the BH formation process.
When the thermodynamical trajectory go below, above, or through CP
in asymmetric matter, the QCD phase transition proceeds via
the first order transition, the cross over transition, or the CP sweep.
When CP is swept,
the density fluctuation should grow,
and we can expect various effects around CP
such as the baryon density fluctuations, focusing of thermodynamical 
trajectories, disappearance of the sound mode, as discussed
for the beam energy scan at RHIC~\cite{Sig-CEP}.

It is a big challenge to construct an EOS which is applicable
to the dynamical simulation of the core-collapse processes,
contains both the hadronic and quark degrees freedom,
and includes CP of QCD.
There is an attempt to include both quark and baryon contributions
based on the PQM~\cite{Steinheimer}.
In that work, the vacuum quark contribution is ignored
and the QCD phase transition becomes the first order
in the two-flavor chiral limit.
It may be also necessary to consider the effects of inhomogeneous
phases~\cite{Inhomogeneous},
which may emerge around the first order phase boundary.

This work was supported in part
by Grant-in-Aid for Scientific Research from JSPS and MEXT
(Nos. 22-3314, 22540296),
the Grant-in-Aid for Scientific Research on Innovative Areas
from MEXT (No. 20105004),
the Yukawa International Program for Quark-hadron Sciences (YIPQS),
and by Grants-in-Aid for the global COE program
`The Next Generation of Physics, Spun from Universality and Emergence'
from MEXT.








\end{document}